\documentclass[preprint]{aastex}
\usepackage{fullpage}

\usepackage{natbib}

\slugcomment{Not to appear in Nonlearned J., 45.}
\shorttitle{Unified Monte Carlo Simulation}
\shortauthors{Chang \& Herbst}

\begin{document}

\title{A Unified Microscopic-Macroscopic Monte Carlo Simulation of Gas-Grain Chemistry in Cold Dense Interstellar Clouds}
\author{Qiang Chang and Eric Herbst\altaffilmark{1}}
\affil{Department of Chemistry, University of Virginia,
              Charlottesville, VA 22904 USA}
\altaffiltext{1}{Also: Departments of Astronomy and Physics,
University of Virginia, Charlottesville, VA 22904 USA}

\begin{abstract}
For the first time, we report a unified microscopic-macroscopic Monte Carlo simulation of gas-grain chemistry in cold interstellar
clouds in which both the gas-phase and the grain surface chemistry are simulated by a stochastic technique. 
The surface chemistry is simulated with a microscopic Monte Carlo method in which the chemistry occurs on an initially flat surface.  
The surface chemical network consists of 29  reactions initiated by the accreting species H, O, C, and CO.  Four different models are 
run with diverse but homogeneous physical conditions including temperature, gas density, and 
diffusion-barrier-to-desorption energy ratio.  As time increases, icy interstellar mantles begin to grow. 
Our approach allows us to determine the morphology of the ice, layer by layer, as a function of time, and to ascertain the 
environment or environments for individual molecules.
Our calculated abundances can be compared with observations of ices and gas-phase species, as well as the results of other models.  

\end{abstract}

\section{Introduction}
Interstellar molecular clouds are uniquely large chemical systems. For instance, 
 giant molecular clouds are typically of dimension 20-100 pc while molecular cores have a size $\approx  0.1$ pc. 
Yet, in these large entities, species such as H$_2$ and ices such as water are mainly formed on surfaces of tiny dust particles with radii ranging in size from less than 0.01$\mu$ to somewhat more than  0.1$\mu$. 
Species are exchanged among numerous independent  dust grains and a large gas-phase body by accretion and desorption processes.  

The extent of a chemical reaction system in terms of density and physical size can determine the mathematical approach by which we calculate the temporal evolution of chemical abundances.  Chemical reaction events are essentially discrete processes, so fluctuations exist for the evolution of chemical species in 
any system.  Because molecular clouds contain large amounts of gaseous material, however, the fluctuations due to randomness of the chemical reaction events
in the gas phase cancel out, so that we can adopt the rate-equation (RE) approach to calculate the evolution of species in the gas phase \citep{GWH2008}.
One might think as well that because there are so many dust particles in clouds, the fluctuations of chemical species on each dust grain can also 
cancel out so rate equations are still applicable. It turns out that this hypothesis is true only when the average number of individual  reactive species per 
 dust grain is not well below 1 \citep{Tielens1997}. When the average number of  reactive species is significantly below unity, significant mistakes can be made with the 
RE approach \citep{Tielens1997, Caselli1998, Herbst2003, Biham2003, Stantcheva2004}. 

Various approaches have been suggested and tested to solve this problem of the surface chemistry on dust particles. The modified rate equation approach is the most computationally 
efficient one and can be implemented in a large chemical reaction network \citep{Caselli1998, Garrod2008,Garrod2009}. 
However, because of its empirical nature, the correctness of this approach cannot be guaranteed in all instances, and should be  
tested by more rigorous approaches, which are stochastic in nature. Solution of the chemical master equation, which describes  the evolution of probability of so-called occupancy states,
is much more rigorous. There are two ways to solve the chemical master equation, and both are computationally tedious.
The so-called ``macroscopic Monte Carlo approach'' was the first  to be applied in astrochemical simulations.  Initially, it was used to 
simulate gas-phase or surface kinetics separately \citep{Charnley1998, Charnley2001}, but eventually   
 was used to simulate the chemistry occurring with a  full gas-grain chemical reaction network \citep{Vasyunin2009, Charnley2009}. 
The advantage of a Monte Carlo simulation is that it actually 
is a numerical experiment that realizes one possible trajectory for the evolution of occupancy states of different species; thus,  one does not
have to consider all possible occupancy states. The chemical master equation can also be directly integrated \citep{Biham2001, Green2001}.
The advantage of direct integration is that it can be easily coupled to RE simulation of gas-phase kinetics.
Because of the large number of possible occupancy states, however, approximations have to be made and so far this method has only been used in small systems \citep{Stantcheva2004}. 
There is another approximate approach, the moment equation method, which, at low order, is much more computationally efficient \citep{Biham2003, Barzel2007}, but is also 
less rigorous.  Recently, it has been successfully applied in a  hybrid approach in which it is turned on and off depending upon the number of species per grain to simulate a large gas-grain chemical reaction network \citep{Du2011}.   The lowest order moment equation approach bears a strong similarity to the most recent modified rate equation approach (A. I. Vasyunin 2012, private communication).   
      
Another problem that occurs in the simulation of gas-grain chemistry derives from the complexity of surfaces. Surfaces can be rough; on such surfaces,  diffusion and desorption rates for species can vary depending on the local position of each particle \citep{Cuppen2005}. 
Although the diffusive, or Langmuir-Hinshelwood, mechanism for surface chemistry is the dominant and best understood process, other mechanisms exist, including the Eley-Rideal process, in which a gas-phase species lands directly on an adsorbate, and the hot-atom mechanism, in which a gas-phase species lands obliquely on a surface and travels in a non-thermal manner for a considerable distance.  
So far,  the only method that can rigorously treat problems with complex surfaces  is the continuous-time random walk (CTRW) Monte Carlo simulation 
approach \citep{Chang2005}, also known variously as a microscopic Monte Carlo or kinetic Monte Carlo approach.    With the microscopic Monte Carlo approach, 
one can keep track of the position of each particle on a surface, which is represented as a planar lattice, so that processes such as diffusion and desorption can be tracked based on  
the local environment of each adsorbate.  The microscopic Monte Carlo simulation was initially used to simulate molecular hydrogen formation on dust 
grains \citep{Chang2005,Cuppen2005}, and is still used for this purpose \citep{Iqbal2012}.  Inclusion of rough or amorphous surfaces extends the temperature range of efficient H$_{2}$ formation at low temperatures, in agreement with experiment \citep{Chang2005,Vidali2009}.  The technique has also been applied to simulate surface chemistry occurring with larger surface reaction networks \citep{Cuppen2009, Cazaux2010}.
Based on previous simulations, it would appear that the complexity of surfaces cannot necessarily be ignored.   

Up to now, most  Monte Carlo simulations of interstellar chemistry have contained the assumption that the abundances of gas-phase species stay constant instead of simulating a full  gas-grain reaction network.  
 It is difficult to couple a microscopic or macroscopic Monte Carlo simulation of surface reactions with an RE simulation of reactions in the gas phase because the RE simulation describes an infinitely large
system without fluctuations while the Monte Carlo simulation describes a finite system with fluctuations.   Mathematically, the Monte Carlo simulation gives the abundances of species as discrete numbers while the abundances of  species by an RE simulation are continuous real numbers.  
Nevertheless, we developed an iterative method 
that combines these two  approaches for the microscopic case \citep{Chang2007}.  The approach is successful only when the ratio of the accretion rate to the desorption rate of  species remains constant or the desorption rate of species from grain surfaces is negligible.  An alternative is to use a Monte Carlo approach for both the surface and gas-phase chemistries.  
 Recently \citet{Vasyunin2009} succeeded in such an approach using a macroscopic Monte Carlo method; they isolated a finite volume of gas phase species surrounding a dust particle and did a full gas-grain Monte 
Carlo simulation. (See also \citet{Charnley2009}.) The agreement with the RE approach depends upon the physical parameters assumed; for conditions in which a stochastic treatment is needed, dramatic differences in abundances can occur.  Closer agreement with the results of a modified rate method for the surface chemistry coupled with an RE treatment for the gas chemistry was subsequently obtained \citep{Garrod2009}.

In this paper, we report the first successful unified microscopic-macroscopic Monte Carlo simulation with a gas-grain reaction network.  In this approach, gas-phase 
kinetics are treated macroscopically while grain-surface kinetics on a rough surface are treated by a microscopic simulation. The gas-grain 
model that we used is introduced in Section 2, while our Monte Carlo technique is introduced in Section 3. Section 4 contains
the results of our simulation, and a comparison with results from other techniques as well as observations is reported in Section 5.  Section 6 contains a discussion along with our conclusions.

\section{Gas-grain Model}
Gas phase species and reactions are taken from the benchmark study of  \citet{Semenov2010}, which contains 4406 reactions and 459 species. 
We choose physical conditions to  pertain to cold dense cloud cores, with a standard density of hydrogen nuclei $n_{\rm H}$ of $2\times 10^4$ cm$^{-3}$, 
and  temperatures between 10 K and 16.5 K because some parameters are available only within this temperature range, as explained later. 
Each model is run at a fixed temperature; we do not consider warm-up and possible ice evaporation due to proximity to protostellar sources of radiation.
We maintain a standard dust-to gas number density ratio of $10^{-12}$, with dust grains of radius 0.1$\mu$. We  vary 
$n_{\rm H}$ 
to calculate the dependence of the abundances and environments of
surface species on this parameter. The visual extinction, $A_{\rm v}$,  is fixed at 10 mag, while the cosmic ray ionization rate $\zeta_{\rm H_2}$ 
is $1.3\times 10^{-17}$ s$^{-1}$.   
Initial gas-phase abundances are given in Table~\ref{table1}; these, taken from \citet{Semenov2010}, refer to 
low-metal abundances except for values based on new measurements of C, N, and O \citep{WH2008}.

\begin{table}
\caption{Initial Gas Phase Abundances}
\label{table1}
\begin{tabular}{lc}
  \hline \hline
  Species & Fractional Abundance w.r.t. $n_{\rm H}$ \\
  \hline
  He       & $9.00 \times 10^{-2}$ \\
  H$_2$    & $5 \times 10^{-1}$ \\
  C$^{+}$  & $1.2 \times 10^{-4}$\\
  N        & $7.6 \times 10^{-5}$\\
  O        & $2.56 \times 10^{-4}$\\
  S$^{+}$  & $8.0 \times 10^{-8}$\\
  Si$^{+}$ & $8.0 \times 10^{-9}$\\
  Na$^{+}$ & $2.0 \times 10^{-9}$\\
  Mg$^{+}$ & $7.0 \times 10^{-9}$\\
  Fe$^{+}$ & $3.0 \times 10^{-9}$\\
  P$^{+}$  & $2.0 \times 10^{-10}$\\
  Cl$^{+}$ & $1.0 \times 10^{-9}$\\
  \hline  
\end{tabular}
\end{table}

We start from a flat surface and choose the density of sites 
to be  $s=1.5\times 10^{15}$ cm$^{-2}$, as measured for olivine \citep{Katz1999}.  Thus there are $1.88 \times10^6$ sites per dust grain. 
We assume that the flat surface consists of unit surface cells such that each site has four nearest neighbors.  As the ice builds up, we assume a simple cubic structure in which each species can have six nearest neighbors, including one above and one below the plane in which the species resides.  The detailed surface model has been 
explained in our earlier paper \citep{Chang2007}.   The desorption energy and diffusion
barrier are the main parameters that determine the mobility of species on grain surfaces.  On a  rough surface, these parameters can vary with position,  so that mobility depends on the local environment.   The values of the desorption energy and diffusion barrier come from two contributions.  
The vertical interaction is that  between the adsorbate and the species lying below it. 
We use $E{\rm _ D}^0$ and $E_ {\rm b}^0$ to represent the vertical desorption energy and vertical diffusion barrier, respectively. 
The other contribution derives from lateral bonds, which are from interactions between the species and their horizontal neighbor molecules. 
Each molecule on a site can form up to to 4 lateral bonds with its neighbor molecules. The actual number of bonds formed is dependent 
on the location of the molecule. The strength of lateral bonds is not known very well.  In order to explain the high efficiency of H$_2$ formation over a wide temperature range, the strength
of lateral bonds cannot be too small in order to  trap enough H atoms on the grain surface \citep{Cuppen2005}.  However, if the lateral bond strength 
is very large, species on a grain surface will lose much of their mobility, and so-called ``islands'' of adsorbates will tend to form, making the surface effectively ``rough.'' 
We choose $E_{\rm iL}$, the lateral bond strength for any species $i$, 
 to be 10\% of its vertical desorption energy, $E_{\rm D}^0$, a minimal value according to \citet{Cuppen2005}, but one that still broadens the temperature range for efficient H$_{2}$ formation.
The local desorption energy is $E_{\rm D} = E_{\rm D}^0 + nE_{\rm iL}$ \citep{Cuppen2005}, where $n$ is the number of lateral 
bonds formed by the species and neighbor molecules. Similarly, $E_{\rm b} = E_{\rm b}^0 + nE_{\rm iL}$.
Table ~\ref{table2} gives $E_{\rm D}^0$ values taken from  \citet{Garrod2006}
which we use for each surface species considered here. 
We choose $\rho=$ $E_{\rm b}^0$/$E_{\rm D}^0$ to be 0.77 or 0.5 
in our simulations, which corresponds to a slow diffusion rate and a fast  rate,  respectively, assuming the better known desorption energy to be fixed.
The high value derives from the experiment on H$_{2}$ formation on olivine by \citet{Katz1999}, whereas the low value is more suitable  for ices \citep{Garrod2006}.
The rate that a species hops from one site to a nearest neighbor site is given by the expression $b_1 = \nu \exp(-E_b/T)$, where $E_b$ is the local diffusion barrier,
$T$ is the grain  temperature and $\nu$ is the trial frequency, which is normally chosen to be $10^{12}$ s$^{-1}$ for physisorbed species.
Similarly, the rate of thermal desorption is expressed by the relation
$b_2 = \nu \exp(-E_{\rm D}/T)$. We assume there is no tunneling under the diffusion barrier because none was found for 
analogous  systems (e.g. olivine) studied in the 
laboratory  \citep{Katz1999}.  We also assume that the products of surface reactions other than H$_2$ do not leave the surface with energy taken from the 
exothermicity of reaction,  although it is possible that they do so according to a statistical mechanism \citep{Garrod2006}.  
Based partially on the analysis of \citet{Katz1999} and partially on the need to minimize the simulation time, however, we allow all newly formed H$_2$ to leave the surface.

\begin{table}
\caption{Vertical Desorption Energies}
\label{table2}
\begin{tabular}{lr}
  \hline \hline
  Species & $E_{\rm D}^0$(K) \\
  \hline
  H$_2$ & 430\\
  H & 450  \\
  O & 800   \\
  CO & 1150 \\
  C &  800 \\
  OH &  2850 \\
  HCO & 1600 \\
  H$_2$CO & 2050 \\
  H$_3$CO & 5080  \\
  CH  & 925  \\
  CH$_2$ & 1050 \\
  CH$_3$ & 1175 \\
  O$_2$ & 1000 \\
  CO$_2$ & 2575 \\
  CH$_3$OH &  5530  \\
  CH$_4$ &  1300 \\
  HO$_2$ &  3650 \\ 
  HOOH   &  5700  \\ 
  H$_2$O  & 5700\\
  \hline
\end{tabular}
\end{table}

Photodesorption is included in our simulation. The flux of FUV photons is given by
\begin{equation}
F_{FUV}=G_0F_0 e^{-\gamma Av} + G_{0}^{'}F_0,
\end{equation}
where $F_0=10^8$ photons cm$^{-2}$ s$^{-1}$ represents the standard interstellar radiation field, while $G_0$ and $G_{0}^{'}$ are scaling factors. We adopt the value for the exponential factor
$\gamma=2$ from the  experimental analysis of \citet{Oberg2007}. The first term in the equation is for external photons while
the second term is for cosmic ray induced photons. Based on a calculation by \citet{Shen2004}, the scaling factor $G_{0}^{'}$ for cosmic ray induced photons
is 4 orders of magnitude smaller than the external radiation scaling factor $G_0=1$.

Table~\ref{table3} shows the chemical reaction network that is used for grain-surface chemistry in our model.  The reactions are of two types: those needed to produce major species such as water, methanol, and methane and used previously,  and those needed to eliminate high abundances of radicals, which can otherwise get trapped in the ice, especially at low temperatures.  The destruction of radicals is assumed to lead to non-radical products, as found in gas-phase databases,  rather than to other radicals.   For example,   OH and CH$_{2}$ do not add to form the radical  CH$_{2}$OH  but produce H$_{2}$CO and H.  Although the addition reactions to form more complex radicals are doubtless present, their exclusion simply allows us to maintain a small number of surface reactions.   

 Chemical activation energy barriers, if non-zero, are also listed and referenced. Because the microscopic Monte Carlo simulation is very computationally demanding, 
we allow only H, O, CO, and C to accrete onto dust grains and to react, starting a chemistry that contains  29 surface reactions.
The sticking coefficient is assumed to be unity.   As the chemistry progresses and ices form on dust particles, the ice  consists mainly  of water, CO, methanol, 
CO$_2$, and CH$_4$. The formation of methanol is particularly interesting, because it is formed by gradual 
addition of H atoms to CO:  CO $\rightarrow$ HCO $\rightarrow$ H$_2$CO $\rightarrow$ H$_3$CO $\rightarrow$ CH$_3$OH. Table~\ref{table4} shows 
the barriers for reactions H + CO $\rightarrow$ HCO and H + H$_2$CO $\rightarrow$ H$_3$CO at different temperatures, as determined experimentally by \citet{Fuchs2009}.  
Such barriers are not as critical for surface reactions as 
for gas-phase processes, because they compete with diffusion barriers \citep{HM2008}.  
The activation energy of the reaction CO + OH $\rightarrow$ CO$_{2}$ + H has not been studied on surfaces. Here we use a value, 176 K,  taken from  the udfa gas-phase reaction network (http://www.udfa.net/). 

\begin{deluxetable}{llrl}
\tablecaption{Surface Reaction Network}
\tablewidth{0pt}
\tablehead{ Number & Reaction & $E_a(K)$ & Reference}
\startdata
1 & H + H $\rightarrow$ H$_2$ &  &\\
2 & H + O $\rightarrow$ OH  & & \\
3 & H + CO $\rightarrow$ HCO & Table 4 & \citet{Fuchs2009}\\
4 & H + C $\rightarrow$ CH & & \\
5 & H + OH $\rightarrow$ H$_2$O & & \\
6 & H + HCO $\rightarrow$ H$_2$CO & &  \\
7 & H + H$_2$CO $\rightarrow$ H$_3$CO & Table 4 & \citet{Fuchs2009} \\
8 & H + H$_3$CO $\rightarrow$ CH$_3$OH & & \\
9 & H + CH $\rightarrow$ CH$_2$ &  &\\
10 & H + CH$_2$ $\rightarrow$ CH$_3$ &  &\\
11 & H + CH$_3$ $\rightarrow$ CH$_4$ & & \\
12 & O + O $\rightarrow$ O$_2$ & & \\
13 & O + OH $\rightarrow$ O$_2$ + H & & \\
14 & O + HCO $\rightarrow$ CO$_2$ + H & \\
15 & O + H$_3$CO $\rightarrow$ H$_2$CO + OH &  &\\
16 & O + CH $\rightarrow$ CO + H &  &\\
17 & O + CH$_2$ $\rightarrow$ CO + H$_2$ & & \\
18 & O + CH$_3$ $\rightarrow$ H$_2$CO + H & & \\
19 & CO + OH $\rightarrow$ CO$_2$  + H & 176 & www.udfa.net\\
20 & C  + OH $\rightarrow$ CO +  H &  &\\
21 & C  + HCO $\rightarrow$ CO + CH & & \\
22 & CH$_3$ +   HCO $\rightarrow$ CH$_4$  +  CO & &  \\
23 & OH  +  H$_2$CO $\rightarrow$ HCO   + H$_2$O &  &\\
24 & OH  +  CH$_2$ $\rightarrow$ H$_2$CO  + H & & \\
25 & C   + O $\rightarrow$ CO & & \\
26 & C + O$_2$ $\rightarrow$ CO + O & \\
27 & H + O$_2$ $\rightarrow$  HO$_2$ & 1200 & \citet{Cuppen2007}   \\
28 & H + HO$_2$ $\rightarrow$  HOOH & \\
29 & H + HOOH $\rightarrow$ H$_2$O + OH & 1400 & \citet{Cuppen2007} \\
\enddata
\label{table3}
\end{deluxetable}


\section{Monte Carlo Method}
Our Monte Carlo simulation method combines the microscopic approach for the surface chemistry 
and the macroscopic approach for the gas-phase chemistry \citep{Gibson2000,Chang2005, Cuppen2005}.   In this
section, we first introduce these two methods and then show how they can be unified.

To perform a microscopic Monte Carlo simulation of chemical reactions on a grain surface with $N$ binding sites, 
we put these sites on a square lattice with dimensions $L \times L$, 
where $L=\sqrt N$. Events are executed in order of the absolute time when these events occur. 
There are three different types of events 
occurring on a lattice. Species accrete from the gas phase and land on  random binding sites on the lattice. 
The absolute time of accretion is calculated by gas-phase kinetics,  which will be discussed below.
A species can either hop from one site to any one of its four nearest horizontal neighbor sites  with equal probability 
or desorb from the surface.  The hopping motion is subject to periodic boundary conditions on the lattice.  We generate a random number $X$ distributed uniformly within (0, 1) to determine whether a given species will perform hopping or thermal desorption. 
If $X<b_1/(b_1+b_2)$, a species will hop, otherwise, it will desorb. The ``waiting time''  before hopping
or desorption happens is calculated as $\tau = -\ln(X^{'})/(b_1+b_2)$, where $X^{'}$ is another random number. 
This expression for $\tau$ maps out a homogeneous Poisson distribution which mimics the unimolecular decay of a species in a given lattice point.
The absolute time is then calculated as $t= t^{'}+ \tau$ where $t^{'}$ is the time when the species first arrived at the binding site.

Chemical reactions are results of these different types of events.  For surface reactions, we assume that a reaction occurs instantly when two species arrive
in  the same site 
 and can undergo a chemical reaction without activation energy. There are three mechanisms included here. The first 
one, known as the Langmuir-Hinshelwood mechanism, occurs  by pure diffusion; i.e., a species hops to a neighboring site already occupied by another adsorbate and the two react.  This mechanism is  the major one studied by surface chemists.  The second one is called the Eley-Rideal mechanism.  Here a gas-phase species lands directly on a site where there is already a reactive adsorbate.  In our simulations, this process is typically a minor one. We label
the third one a ``chain reaction'' mechanism, which will be  explained later. 
If there is activation energy, we simulate a competition between diffusion away from the site and tunneling under or
hopping over the activation energy barrier to lead to reaction for both the Langmuir-Hinshelwood and Eley-Rideal mechanisms \citep{HM2008}.  
The activation energy potential is modeled as a square potential with width $a$ = 1~\AA, 
so the rate of  tunneling  is calculated as $b_3 = \nu \exp(-\frac{2a}{\hbar}\sqrt{2\mu E_a})$, 
where $E_a$ is the reaction barrier, $\nu$ is the trial frequency, and $\mu$ is the reduced mass of reactants.  
Once we calculate $b_3$,  
a random number $Y$ uniformly distributed within (0, 1) is
generated to determine if the reaction can happen via tunneling.  If $Y<b_3/(b_1+b_2+b_3)$, the reaction will happen, otherwise, the reaction will not happen, 
where we have also included the rate of thermal desorption as a competitive process.
In addition to surface reactions, we allow H atoms to diffuse into the ice.
When an H atom comes to a site by hopping or accretion, we determine if there is any species in the top 4 monolayers on the site that can 
react with H atoms without a chemical barrier.  If there is such a species, a reaction without barrier happens with the topmost species within the top 4 monolayers. 
Otherwise, we check to see if there is any species in the top 2 layers on the site that can react with H atoms via an activation barrier. 
If there is such a species, we use the competition mechanism to determine if the reaction with a barrier 
can happen or not with the topmost species.

\begin{table}
\caption{Activation Energy Barriers}
\label{table4}
\begin{tabular}{ccc}
  \hline \hline
T (K) & CO + H & H$_2$CO + H \\ \hline
12.0 & 390 & 415 \\
13.5 & 435 & 435 \\
15.0 & 480 & 470 \\
16.5 & 520 & 490 \\
\hline
\tablecomments{ The activation barriers are from~\citet{Fuchs2009}. 
}
\end{tabular}
\end{table}


Some surface reactions are handled in a distinctive manner.  For the two surface reactions in Table~\ref{table4}, \citet{Fuchs2009} 
fit their experimental data to  the hopping expression $b_3 = \nu \exp(-E_a/T)$ with a temperature-dependent activation energy, 
and a lower trial frequency of $2 \times 10^{11}$~s$^{-1}$. 
We use this frequency and these values of the activation energy  in our simulation in order to match the experimental fitting procedure, where the lower $\nu$ value is also used to calculate the  competitive diffusion and desorption rates.

We also must consider a phenomenon we describe loosely as a  ``chain'' reaction.  Suppose that there is a site with O on the topmost layer and CO underneath it. 
The reaction between these two species is not included in our small surface network because it has a sizable barrier.  
If a surface H atom diffuses onto the outermost layer of the site, it can react quickly with O to form the radical OH, 
which in turn can react with the CO molecule to form CO$_{2}$ and H despite a small activation energy. This unusual type of ``chain'' reaction has already been
implemented  with the modified rate method in a three-phase gas-grain model, where the so-called ``third'' phase is the outermost layer \citep{Garrod2011}. Here, because of the spatial resolution
of the microscopic Monte Carlo simulation, we are able to treat this type of unusual reaction more rigorously. 

The macroscopic Monte Carlo approach is used to simulate the gas-phase chemistry \citep{Gibson2000,Vasyunin2009,Charnley2009}. 
In this approach,  spatial information within a volume of space $V$ surrounding a dust particle used in the simulation is
not obtained.  We start with a number of molecules for each species based on initial abundances and the volume of space. 
We then calculate the reaction rate $r$ (s$^{-1}$) for each reaction in the volume $V$.
The waiting time for reaction $i$ is determined by $\tau_i = -\ln(Z)/r_i$ where $Z$ is yet another random number uniformly 
distributed over (0, 1). The absolute time
when the i-th reaction happens is $t_i = t_{i0} + \tau_i$, where $t_{i0}$ is either the time when the i-th 
reaction happened  previously or 0, when we start the simulation. 
We sort the absolute time for each reaction and 
find the smallest time at each step.   If the i-th reaction happens first, at a  time $t_i$, 
we update the numbers of all species involved in this reaction. We update the rate of the k-th reaction only 
if the number of its reactant molecules change due to the i-th reaction.  If the reaction rate for the k-th reaction is $r_k$ 
before the i-th reaction happens and  $r_k^{'}$ afterwards, the absolute time when the k-th reaction will occur again is updated to be 
\begin{equation}
\label{euq1}
t_{k}^{'} = t_i + \frac{r_k}{r_k^{'}}(t_{k}-t_i).
\end{equation}
Once again, we find the reaction that will occur next, and continue the procedure.  This particular method is, not surprisingly, known as the next reaction approach.

The rate for the i-th gas phase reaction, where the word ``reaction'' includes accretion onto a grain, is calculated in one of three ways, as shown by the relation
\begin{equation}
r_i = \left\{
\begin{array}{ll}
k_i N_j &\mathrm{ one-body~reactions}\\
k_i N_k N_l &\mathrm{two-body~reactions}\\
\frac{N_j L^2}{Vs}\sqrt{\frac{8 k_B T_g}{\pi m_i}} &\mathrm{accretion}\\
\end{array}
\right .
\end{equation}
where $N_j$, $N_k$, $N_l$ are numbers of assorted reactants, $k_i$ is a rate coefficient, $T_g$ is the gas phase temperature, 
$L$ is the dimension of the grain lattice ($N^{1/2}$),  $s$ is the density of sites, and $m_i$ is the mass of an accreting species. 
The next reaction method essentially models each chemical reaction as an inhomogeneous Poisson Process, 
with a rate that can change as other species change in the simulation.  The fact that the rate is not a constant 
makes the merger of the two different Monte Carlo simulations straightforward, since desorption from the ice into the gas 
is just another process leading to a change of rate in the gas phase. 

Photon arrival events are modeled as independent homogeneous Poisson processes.   We posit that only species in the top two monolayers can be photodesorbed, so the
photodesorption rate is initially assumed to be
\begin{equation}
k_{pd}=2F_{FUV}Y_{pd}\pi r_{d}^2,
\end{equation}
where we assume $Y_{pd}$ to be 0.001 for all species, a number that is supported by experimental data~\citep{Oberg2009a, Oberg2009b}. 
The waiting time for a photodesorption is 
$t_{pd}=-\ln(W)/k_{pd}$ where $W$ is a random number uniformly distributed within (0, 1).  Once a photodesorption event can happen, we randomly assign
a site where it might take place.  If the site is empty, then the photodesorption is a null event. If there is only one layer
of species on that site, we generate yet another random number $U$ to determine if photodesorption events can happen or not.  If $U\leq0.5$, 
a photodesorption event occurs, otherwise, this is a null event. If there are at least two monolayers of species on that site, we randomly pick
one of the species in the top two monolayers and allow this species to desorb into the gas phase. For simplicity, we ignore photodissociation
in our reaction network.

The unification of the microscopic and macroscopic Monte Carlo procedures is straightforward.   Each event in the gas-grain system has an absolute time when it happens. 
We sort out the smallest one and execute that event, which could be hopping, hopping leading to reaction, accretion, 
desorption, or reaction in the gas phase.  Accretion is considered to be  
a gas phase reaction. Once an accretion event  happens, we update the number of accreting species, all necessary 
gas-phase reaction rates, and the absolute time of relevant reactions in the gas phase. We then add a molecule of the accreting species to the lattice. 
For a thermal desorption event, we delete the molecule from the lattice and add one molecule of that species to the gas phase. 
We once again update all necessary values in the gas phase.   For all events 
other than desorption or accretion, their execution occurs either on the dust particle or in the gas.    

Surface processes in our simulation require more random numbers than needed with Gillespie's original method~\citep{Gillespie1976}, mainly because of
the extra spatial resolution capability of the microscopic Monte Carlo simulation. For instance, for hopping and evaporation processes, as in Gillespie's 
original method, we use one random number to determine the waiting time and then use the second random number to determine whether to hop or evaporate.
However, in order to find the position of the particle after hopping, we need another random number to determine its location.
On the other hand, it is not necessary to use the third random number for evaporation processes because we do not have to keep track the
position of the evaporating particles.  Similarly, accretion and photodesorption all require
more random numbers  to find the location when these processes happen. 
Moreover, reactions with barriers are modeled as the results of more than one encounter  in the microscopic Monte Carlo simulation, 
so we need extra random numbers to determine if an encounter with a barrier between the two particles can lead to a reaction or not.

\begin{figure}
\centering
\resizebox{10cm}{10cm}{\includegraphics{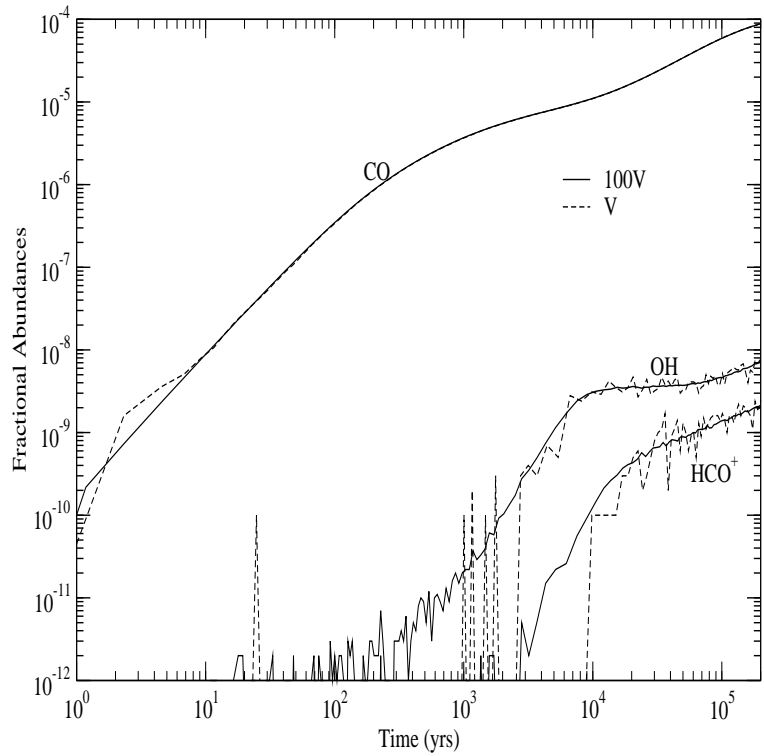}}
\caption{Selected gas-phase abundances using a purely gas-phase Monte Carlo simulation with 
         gas phase volumes 100 $V$ and $V$ at 10~K. The density of H nuclei is fixed
         at $2\times 10^4$ cm$^{-3}$.  
}
\label{fig1}
\end{figure}

Because the microscopic Monte Carlo approach is very computationally expensive, we cannot use a whole standard grain with $1.88\times 10^6$ sites. 
On the other hand, previous calculations show that there is hardly any size dependence for chemical kinetics on a rough grain surface \citep{Cuppen2009}.
So we only take 1\% of a standard grain surface; i.e. $1.88\times 10^4$ sites, consisting of a $137\times137$ lattice.  
The gaseous volume associated with this part of the grain surface is $V= 10^{10}/n_H$ if the dust-to-gas number ratio is $10^{-12}$. 
Figure~\ref{fig1} shows that other than increased fluctuations, calculated abundances for selected gas phase species by a pure gas-phase Monte Carlo simulation 
do not vary with the volume of isolated gas down to a volume of $V$.  
The fluctuations shown in Figure~\ref{fig1} are at a maximum when HCO$^{+}$ and OH have fractional abundances of $\le 10^{-10}$, 
which means that there is at most one molecule of each species in the volume $V$, so that the fluctuations are expected to be large.   
Species of higher abundance show smaller to negligible  fluctuations. To improve matters for species of low abundance, we can always average gas abundances, 
as discussed in Section 4.  
So gas phase chemistry can be simulated within 1\% of the volume associated with 1\% of the grain without introducing any major error.  Because 1\% of 
a grain surface area is large enough to simulate the surface chemistry accurately, 
we can simulate the gas-grain chemistry using 1\% of a grain surface area and an associated volume of gaseous species. 

 We implemented our method using the C language and ran simulations on a Dell Precision T7500n Work station. The CPU running-time varies significantly depending on physical parameters chosen. 
For Model 1 (see Table~\ref{table5p}) at 10 K, it takes only a few minutes to finish a simulation through $2 \times 10^{5}$ yr while for Model 3, it takes about 4 hours to run an analogous simulation. Moreover,
most of the CPU running-time is spent on late times of evolution in the system. This occurs because at these times
H atoms have to hop many times to find partners with which to react. 

\section{Results}

In this section we report the molecular abundances in the ices as they evolve on dust grains at different temperatures, gas-phase number densities, 
and ratios of diffusion barrier to desorption energy, $\rho$. To account for uncertain  parameters,
we used four different models,  shown in Table~\ref{table5p}, and defined by the value of the hydrogen nuclear density $n_{\rm H}$, which is either $ 2 \times 10^{4}$ cm$^{3}$ s$^{-1}$ or $ 1 \times 10^{5}$ cm$^{3}$ s$^{-1}$, and the value of $\rho$, which is either 0.77 or 0.5.   The temperatures at which the models are run
were chosen to be 10~K, 13.5~K, 15~K, and 16.5~K, because the experimental reaction barriers for H + CO and H + H$_2$CO are only available 
at 12 K, 13.5 K, 15 K and 16.5 K \citep{Fuchs2009}. 
Because 10 K is a standardly used temperature, we extended the reaction barriers to 10 K by assuming
that the reaction rates at 10 K are the same as at 12~K as  in \citet{Cuppen2009}.
The dust-to-gas number ratio is fixed  at  $10^{-12}$.
We ran each simulation for a time period of $2\times10^5$ yr, the so-called early time for cold cores, 
where best agreement for gas-phase species is typically obtained. Moreover, the amount of water ice produced at this time is sufficient to explain observations,  as discussed below \citep{Gibb2004,Pontoppidan2004,Boogert2004}. For each
set of parameters, we ran the simulation four times with different random seeds and then calculated the average abundances in order to reduce 
the fluctuation in the results.


\begin{table}
\caption{Parameters for Different Models } 
\label{table5p}
\begin{tabular}{ccccc}
  \hline \hline 
Parameter  & Model 1  & Model 2 & Model 3 & Model 4  \\ \hline
$n_{\rm H}$ (cm$^{-3}$)    & 2.0(4)   & 1.0(5)  &  2.0(4)  & 1.0(5)   \\
$E_{\rm b0}$/$E_{\rm D0}$    & 0.77   & 0.77  &  0.5  & 0.5    \\
\end{tabular}
\tablecomments{ a(b) means a$\times 10^{b}$. 
}
\end{table}

\subsection{Ice abundances at different temperatures and hydrogen densities} 

We first discuss results for Models 1 and 2, where $\rho$ is fixed at 0.77.  For Model 1, the density of hydrogen nuclei is $2\times10^4$ cm$^{-3}$ while for Model 2, it is $10^5$ cm$^{-3}$.  Below we focus on the molecules in our models with major ice abundances over some temperature range.  We start, however, with a look at the  gas-phase species that accrete onto grains.

\begin{figure}
\centering
\resizebox{10cm}{10cm}{\includegraphics{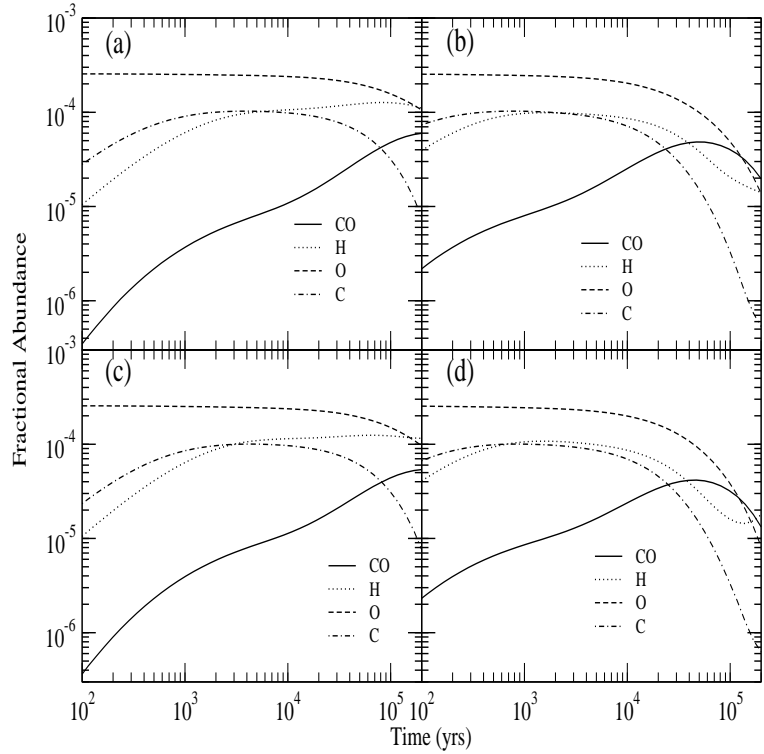}}
\caption{The evolution of the abundances of gas-phase species that accrete onto grain surfaces as followed with Models 1 and 2. 
       Panel (a): $T=10$~K, Model 1,
 panel (b): $T=10$~K, Model 2, panel (c): $T=15$~K, Model 1, panel (d): $T=15$~K, Model 2. 
}
\label{fig2}
\end{figure}

Figure~\ref{fig2} shows the gas-phase fractional abundance vs time of the four accreting species in our models  at two different temperatures (10 K and 15 K) and $n_{\rm H}$ values. 
Over all, we can see that the change of temperature does 
not cause much difference in the abundances of the gas phase species  while the change in hydrogen density does.   Specifically, increasing n$_{\rm H}$ to $10^{5}$ cm$^{-3}$ reduces the abundances of all accreting species due to the larger accretion rate,  especially at the latest times.  Not all species decrease in abundance with increasing time; for example, gas-phase H and CO increase in abundance  over the entire range
of time for $n_{\rm H}$ = 2$\times 10^{4}$ cm$^{-3}$, while C first increases and then decreases at both densities, reflecting the well-known conversion of C$^{+}$ into CO through neutral atomic carbon.

\begin{figure}
\centering
\resizebox{12cm}{12cm}{\includegraphics{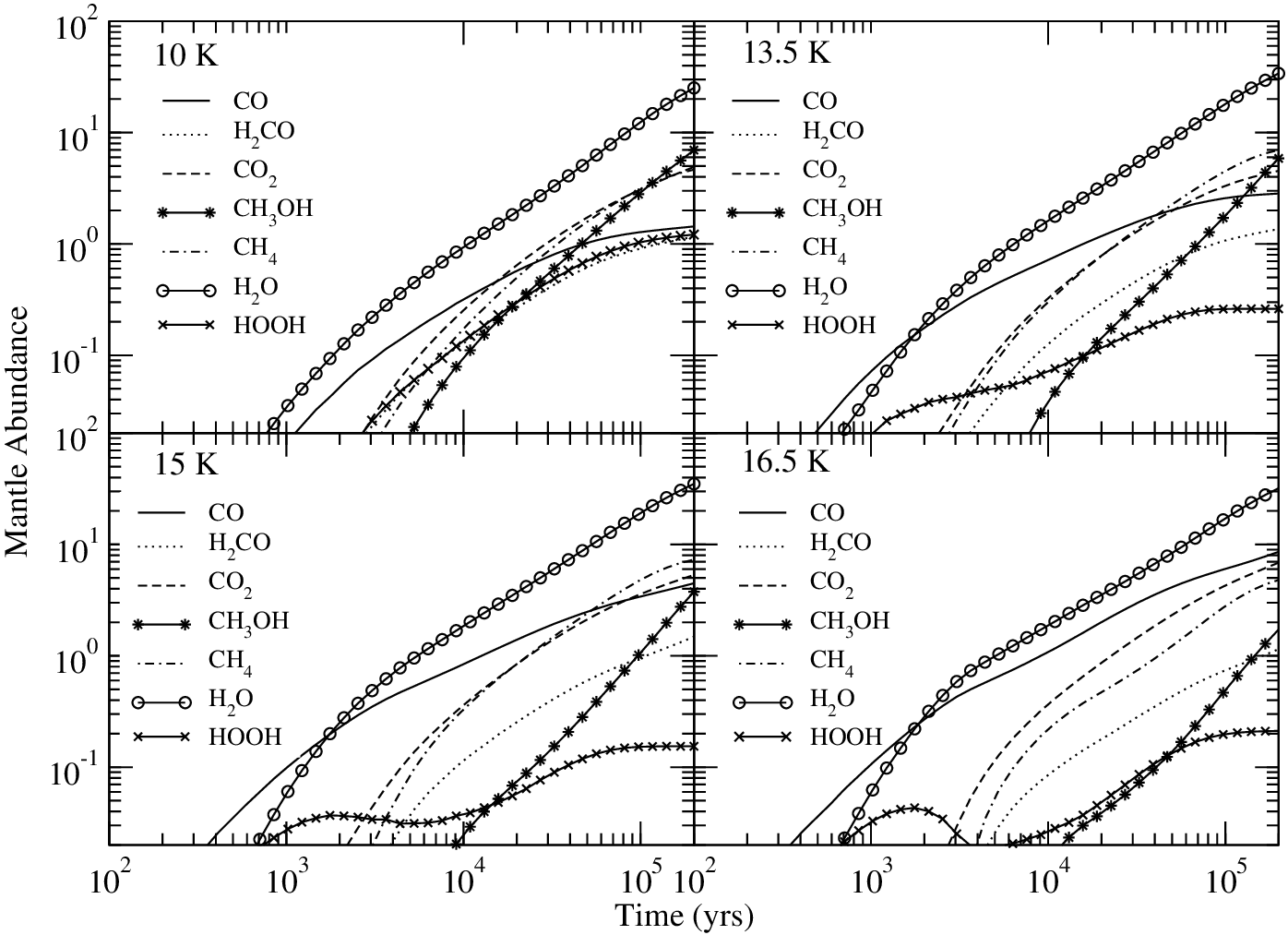}}
\caption{The abundances of water, hydrogen peroxide, and major carbon-containing species in the ice mantle are shown in terms of monolayers as a function of time at different temperatures using Model 1. 
}
\label{fig3}
\end{figure}

Figure~\ref{fig3} shows the build-up with time of CO, H$_2$CO, CO$_2$, CH$_3$OH, CH$_4$,  HOOH, and H$_{2}$O
 in the ice using Model 1 at different temperatures. The molecular abundances on the grain mantle are depicted in units of monolayers, so that  if we have 1 monolayer of CO in the ice, there are $1.88\times 10^4$ CO molecules, although they need not be in the same monolayer. Because the surface we used in the simulation
is 1\% of an actual grain surface for standard-sized grains, there are in reality $1.88\times 10^6$ CO molecules present. Our dust-to-gas number ratio is fixed at $10^{-12}$, so that
the total number H atoms associated with 1 grain is $10^{12}$; thus, the fractional abundance of CO molecules on a grain surface is $1.88\times 10^{-6}$.  Another way of expressing the abundance of ice molecules is to relate them as percentages of  the abundance of water, which nears 50 monolayers toward the end of our calculation.

As can be seen in Figure~\ref{fig3},  almost all of the major ice species other than hydrogen peroxide at early times
increase in mantle abundance with time, some more sharply than others.  
If we look first at CO, we see that with increasing temperature, the CO abundance increases more strongly while the CH$_3$OH abundance behaves in the 
opposite manner.  By 16.5 K, CO is a major component, occupying almost 10 monolayers at the end of the simulation, while there are only 1-2 monolayers 
of CH$_3$OH. At 10 K, on the other hand, the abundances are reversed.  This result is not surprising because an increase in the surface  temperature 
increases the rate of desorption of  hydrogen atoms from the grain surface, thus reducing the H addition rate to CO.  This effect is more important than an increase in rate due to hopping over the H + CO activation energy barrier.  From Table~\ref{table4}, we can see that because the activation energy rises from 390 K to 520 K as the temperature goes from 12.0 K to 16.5 K,  the Boltzmann factor rises only slightly from $7.7 \times 10^{-15}$ to $2.1 \times 10^{-14}$.    The analogous change in the Boltzmann factor for the desorption rate of H is much greater since the desorption energy is constant. A similar statement can be made about the relative lack of temperature dependence for the hopping over the barrier for H + H$_{2}$CO.  Note that the abundance of formaldehyde shows a weak temperature dependence because both H + HCO and H + H$_{2}$CO are reduced in rate due to the increasing desorption of H with increasing temperature.  
Our model can produce a moderate amount of H$_2$CO between 10 K
and 16.5 K. 

\begin{figure}
\centering
\resizebox{12cm}{12cm}{\includegraphics{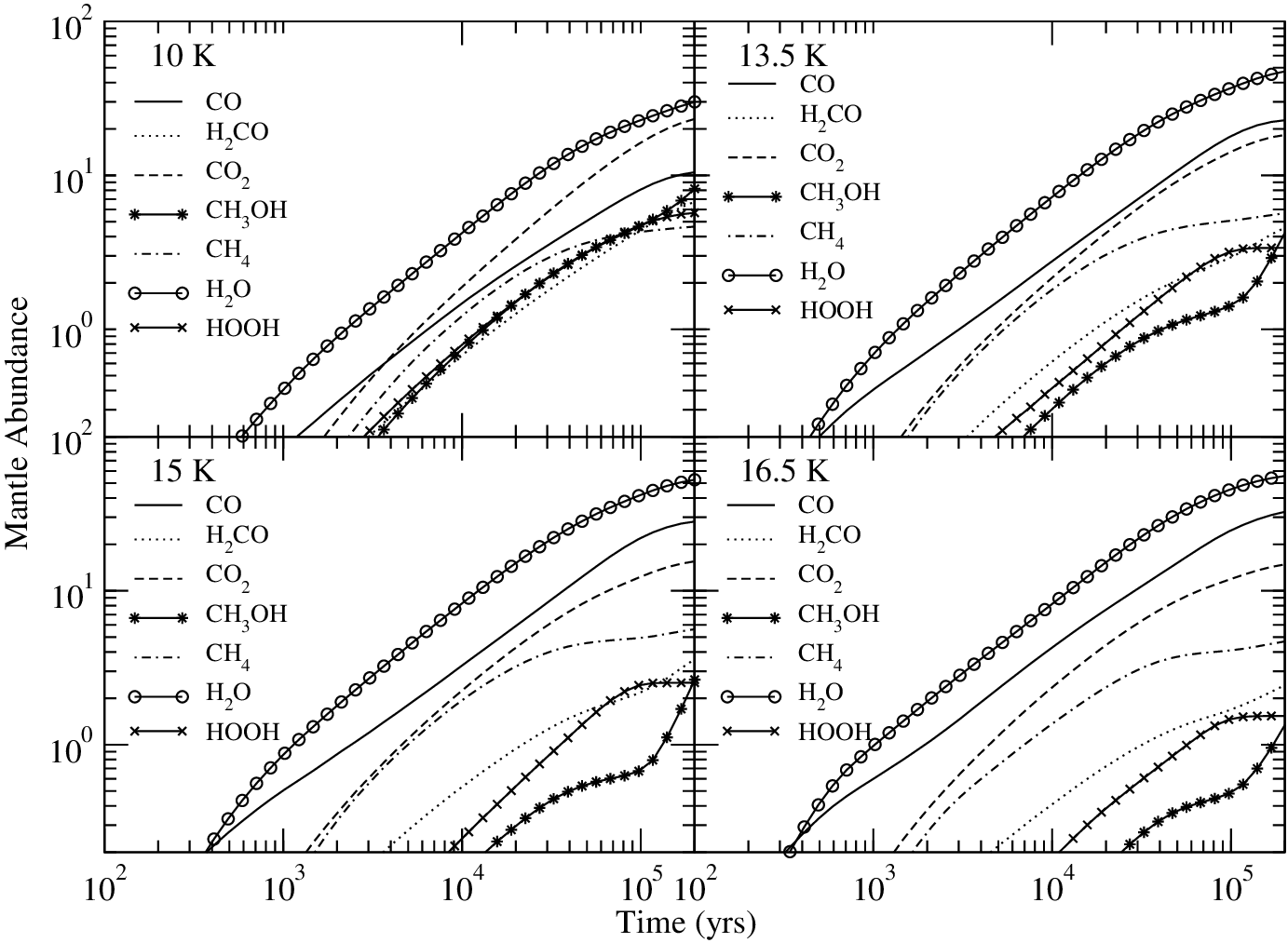}}
\caption{The abundances of water, hydrogen peroxide and major carbon-containing species in the ice mantle are shown in terms of monolayers as a function of time at different temperatures using Model 2. 
}
\label{fig4}
\end{figure}

  Model 1 produces a  significant amount of 
CH$_4$  at all temperatures. The desorption energies of C, CH, CH$_2$, and CH$_3$ are sufficiently high that these species will not desorb rapidly
through 16.5 K, but will be converted to CH$_4$ with high efficiency.  The major alternative is reaction with O atoms to form CO.  These processes are not the dominant ones for CO except at very early times, after which accretion of CO from the gas dominates.
Carbon dioxide is a major carbon-bearing 
molecule that can be formed between two heavy species on grains between 10~K and 16.5~K. This molecule is produced mainly by the reaction 
OH + CO $\rightarrow$ CO$_2$ + H, which occurs by an unusual ``chain mechanism'' discussed earlier, in which an H atom first combines with an O atom lying 
above a CO molecule, so that the OH need not undergo horizontal diffusion to react with CO.  Water, the dominant species, is formed by a variety of surface reaction sequences starting with atomic O, molecular O$_{2}$, and even O$_{3}$.  Use of only the first two syntheses (see Table~\ref{table3}) is sufficient in the temperature range studied to produce  30 monolayers in $2 \times 10^{5}$ yr, far more than could be produced by accretion of gas-phase water \citep{Gibb2004,Pontoppidan2004,Boogert2004}.  The O + H mechanism is dominant at 10 K while O$_2$ + H is the dominant mechanism at 16.5 K.
There are also moderate amounts of HOOH produced on grain surfaces in Model 1. The production of hydrogen peroxide 
generally decreases with higher temperature due to an increase in rate for the reaction
H + HOOH $\rightarrow$ OH + H$_2$O, which possesses a barrier.
We can also see that the abundance of HOOH has a bump at 16.5 K at early times.
This can be explained by our initial abundances and  the rapid increase of H atoms in the gas phase at early times. 
 

Analogous results with Model 2 are shown in Figure~\ref{fig4}, which indicates the influence of higher density.  
One obvious consequence of the increased density is the increased total thickness of ice on the grain surface, 
which occurs because of the increased deposition flux. For example, at temperatures above 10 K, 50 monolayers of 
water ice can be produced in $2 \times 10^{5}$~yr.  The other
major difference is that greater amounts of CO  and H$_2$CO are produced on grain surfaces,  especially at 10 K. 
This increase in the mantle CO abundance occurs for two reasons. First,  the fractional abundance of gas-phase atomic H in the higher density case decreases
strongly after $2\times 10^4$ yr,  as shown in Figure~\ref{fig2},  such that it is far lower than that of CO.   The lowered relative flux of H onto grains compared with CO makes it more difficult to destroy CO.  
Second, CO molecules are more easily buried before H atoms can react with them with the increased flux of species. 

The increased abundance of H$_2$CO is more difficult to understand
because it is more difficult to convert CO to H$_2$CO; however, it is also more difficult to convert H$_2$CO 
to CH$_3$OH at higher density. Based on our simulation, the absolute abundance of H$_2$CO increases, but the abundance 
ratio of H$_2$CO to  CO does not increase at higher density.  The increased abundance of HOOH can be understood in similar way; i.e.,
the rate of the formation reaction (H + HO$_2$) decreases less rapidly than the rate of the destruction reaction (H + HOOH).
From Figure~\ref{fig4}, we can see that  the absolute abundance of CH$_3$OH is somewhat decreased at higher density, while the abundance ratio of 
CH$_3$OH to CO clearly drops precipitously, as explained earlier.


\begin{table}
\caption{Model 1 Abundances of Major Mantle Species} 
\label{table6p}
\begin{tabular}{ccccc}
  \hline 
Species  & 10 K  & 13.5 K & 15 K & 16.5 K  \\ \hline
CO       & 5.7   & 8.4  &  12.8  & 26.8   \\
H$_2$CO  & 4.6   & 4.0  &  4.2  & 3.6    \\
CO$_2$   & 19.6   & 13.4  &  15.2 & 21.4  \\
CH$_3$OH & 27.7   & 17.3  &  10.7  & 5.4 \\
CH$_4$   & 18.5   & 20.6  &  20.9  & 15.0  \\
HOOH     & 4.8    & 0.77  &  0.44  & 0.66 \\ 
H$_2$O   & 4.7(-5)& 6.4(-5) & 6.6(-5) & 6.0(-5) \\ \hline
\end{tabular}
\tablecomments{The abundances pertain to a time of $2\times 10^5$~yr.}
\tablecomments{The abundance of water ice is the fractional abundance with respect to  gas-phase $n_{\rm H}$, while the abundances of the other ice components are percentages with respect to the water ice. a(-b) means a$\times 10^{-b}$. 
}
\end{table}

\begin{table}
\tabletypesize{\scriptsize}
\caption{Model 2 Abundances of Major Mantle Species}
\label{table7p}
\begin{tabular}{ccccc}
  \hline
Species & 10 K& 13.5 K& 15 K & 16.5 K  \\ \hline
CO      & 34.9 & 48.1  & 53.5 & 58.5   \\
H$_2$CO & 22.4 & 9.7  & 6.8  &  4.4  \\
CO$_2$  & 77.3 & 38.7  &  29.5 & 26.6   \\
CH$_3$OH& 27.2 & 8.9  & 5.0   &  2.4 \\
CH$_4$  & 15.5 & 12.0  & 10.7   &  8.4  \\
HOOH    & 19.0 & 7.1  & 4.8   & 2.8   \\
H$_2$O  & 5.6(-5)& 8.9(-5) & 9.9(-5) & 1.0(-4)  \\ \hline
\end{tabular}
\tablecomments{The abundances pertain to a time of $2\times 10^5$~yr.}
\tablecomments{The abundance of water ice is the fractional abundance with respect to gas-phase $n_{\rm H}$, while the abundances of the other ice components are percentages with respect to the water ice.  a(-b) means a$\times 10^{-b}$. }
\end{table}

Table~\ref{table6p} and Table~\ref{table7p} show the abundances of water ice,  hydrogen peroxide, and all major carbon-bearing species 
in the mantle ice at the final time of $2\times10^5$ yr.
Table~\ref{table6p} lists results for Model 1,  while Table~\ref{table7p} lists results for Model 2. 
The abundance of water is given as its fractional abundance with respect to the gas-phase nuclear hydrogen density $n_{\rm H}$, 
while the abundances of the other ice species are given as the percentages relative to water ice.  Some large differences can be seen between results at the two different densities, especially at low temperatures.  For example, the abundance of CO with respect to water is a factor of 6 lower at low density at 10 K, and only a factor of 2.2 lower at 16.5 K.  These results will be compared with observations in Section~\ref{comp}.

Now let us consider the distribution of assorted species throughout the grain mantle at $2\times 10^5$ yr.  
Note that the inner layers were formed at earlier times and remain relatively passive.  
Figure~\ref{fig5} shows the fraction of each monolayer occupied by carbon-bearing species, hydrogen peroxide,  and water
starting on the surface of the grain and ending at the uppermost and partially occupied layers for Models 1 and 2 at 10 K and 15 K.
Because the gas-phase atomic C fractional abundance peaks at early stages, its accretion onto dust particles and conversion to methane by 
successive reactions with atomic hydrogen means that the abundance of
CH$_4$ should peak in the inner layers of the ice. Indeed, in Figure~\ref{fig5}, CH$_4$ generally occupies follows this predicted trend, except for a secondary peak at higher monolayers at 15 K for the  higher density. This anomaly can be explained by the depletion of O, which is a competitor  of atomic hydrogen in reacting with C and carbon hydrides.  

The abundance of CO is typically relatively flat as a function of monolayer until the outermost layers, but actually peaks strongly in the innermost layers for  Model 1 at 10 K.  
This result is non-intuitive because CO normally forms on grains by accretion from the gas, and its production in the gas takes some time \citep{Oberg2011}. There are two explanations for a relatively large amount of CO at inner monolayers in our simulation.   First, CO can be produced on grain surfaces by the reaction C + O $\rightarrow$ CO. 
There are high abundances of  gas phase O and C at very early times of our simulation; thus, upon accretion of these atoms onto the ice, a
significant amount of CO can be produced. Secondly, the abundance of gas phase H atoms is small at early stages, so the flux  of H atoms accreting onto the grain surface is small, which reduces a standard destruction route for CO ice.
To confirm this hypothesis, we increased the initial fractional abundance of atomic H to $4\times 10^{-4}$,
and found that the CO abundance at the inner layers virtually disappears.   
At higher densities,  more CO is produced  at later times by accretion, while at 15 K and lower density,  it is harder to destroy CO with H atoms due to the increase in desorption.   The CO abundance distribution  is also discussed in Section 5, where we compare our results with observations and inferences.


The methanol abundance typically peaks at late stages, especially when the gas phase CO is large. The rapid increase
of methanol at outer layers at 15 K for higher densities is particularly interesting. This effect can be explained by the heavy depletion of gas phase O under these conditions, which leads to a lower accretion flux, so that H atoms on the grain are more likely to convert CO into methanol than to react with surface O.

The abundances of CO$_2$ and water at different layers are close to constant except towards the outermost layers. 
The abundance of H$_2$CO at different layers behaves roughly in the same manner as CO regardless of $n_{\rm H}$. 
 The distribution of hydrogen peroxide is  dependent on the gas phase density. At lower densities, HOOH drops quickly from inner to  outer layers.  The effect  at the outer layers 
is explained by the rapid increase of the gas phase ratio of H to O,  and the fact that HOOH can be converted to 
water efficiently in our temperature range.  At higher densities, HOOH maintains a moderate abundance 
at the inner layers before  dropping  rapidly at the outer layers.  The effect at the inner layers
can be understood by a drop of gas phase H  and the fact that HOOH can be buried more easily at higher densities.

Finally, we show a picture of the 
icy surface for Model 1 at  $2\times10^5$ yr at 10 K in Figure~\ref{fig6}. The figure shows the number of layers at each site on the grain surface.  
We can see that the surface has actually grown to be very rough, so that our explanations should be taken with the proverbial grain of salt.

\subsection{Influence of the ratio of diffusion barrier to desorption energy}
In this subsection, we discuss results when $\rho$ is set to a lowered value of 0.5, which increases the diffusion rate for a fixed desorption energy. 
We define Model 3 and Model 4 in analogy with Models 1 and 2 except for the lower value of $\rho$  (see Table~\ref{table5p}).
It is tedious to finish a microscopic Monte Carlo simulation with the faster diffusion rate at temperatures 
higher than 10 K, so we only give results at 10 K.

The temporal evolution of  accreting gas phase species is very much like that in Figure~\ref{fig2}, as expected. 
Figure~\ref{fig7} shows the temporal evolution of water,   hydrogen peroxide,
and major carbon-bearing ice species on dust particles with Models 3 and 4,  while Figure~\ref{fig8} shows
the fraction of each monolayer occupied by these species at the final time of $2 \times 10^{5}$ yr for these models. Compared with results from the slower diffusion rate ($\rho = 0.77$) at the same
temperature, one major difference is that more CO and 
less CH$_3$OH are produced.  The increase in the rate of diffusion affects the competition between diffusion and reaction for Langmuir-Hinshelwood-type reactions with chemical activation energy barriers, for example H + CO and H + H$_{2}$CO.   If the diffusion
rate increases, it is more likely that H atoms will diffuse away from a site than react. On the other hand, faster diffusion can increase
the probability that an H arrives to encounter a CO molecule at a site. Under the conditions of the simulation, the former effect appears to dominate.  
Table ~\ref{table8} shows the abundances of all major species on a dust grain at $2\times 10^5$ yr compared with the water abundance, as in Tables~\ref{table6p} and \ref{table7p}. 
We see that with the faster diffusion rate, increasing $n_{\rm H}$ can  strongly increase the percentage of CO molecules
on the grain surface as also occurs with the slower diffusion rate. If we consider the total absolute abundance at 10 K of the major carbon-bearing species, we see that raising the diffusion rate does not change this value because the decrease in total abundance with respect to water ice is balanced by an increase in the water ice abundance (see Table~\ref{table8} as compared with Tables~\ref{table6p}  and \ref{table7p}). 
With a faster diffusion rate, the temporal evolution of HOOH and the distribution of HOOH at different layers 
behave similarly to the $\rho = 0.77$ results at higher temperatures, which is reasonable because the major effect of 
increasing the temperature is to increase the diffusion rate of species.  Such an analysis is useful for other species as well.

\section{Comparison with Observation and Other Models}
\label{comp}
The primary purpose of this paper is to introduce a stochastic method to merge two different types of Monte Carlo simulations.   
In doing so, we were forced to confine ourselves to a rather small surface reaction network with only 29 reactions, which is mainly useful for calculating abundances and environments of water ice, hydrogen peroxide, and the major carbon-containing ices.
Nevertheless,  it is still interesting to compare our calculated abundances for major ice species with recent observations and results from other surface models.
Comparison with gas-phase species shows that for the time interval calculated, very little difference exists between the results here and results of gas-phase models.
For comparison with observed ice abundances, we utilize our results at a time of $2 \times 10^{5}$ yr.

\begin{figure}
\centering
\resizebox{10cm}{10cm}{\includegraphics{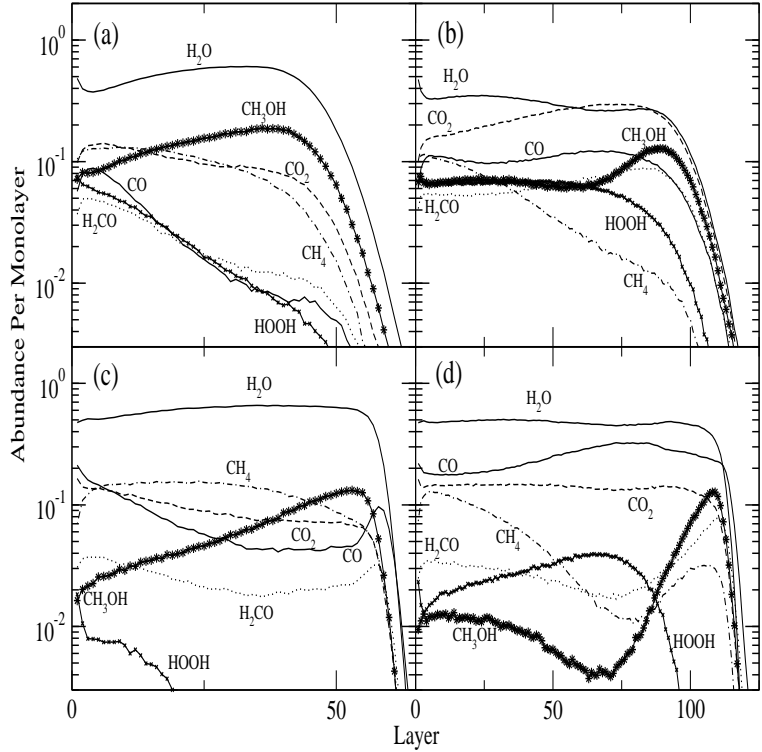}}
\caption{The fraction of each monolayer occupied by  water,  hydrogen peroxide,
and major carbon-bearing species using Models 1 and 2 shown as a function of monolayer at the final time of $2 \times 10^{5}$~yr.
Panel (a): $T=10$~K, Model 1,
 panel (b): $T=10$~K, Model 2, panel (c): $T=15$~K, Model 1, panel (d): $T=15$~K, Model 2. }
\label{fig5}
\end{figure}

\begin{figure}
\centering
\resizebox{10cm}{10cm}{\includegraphics{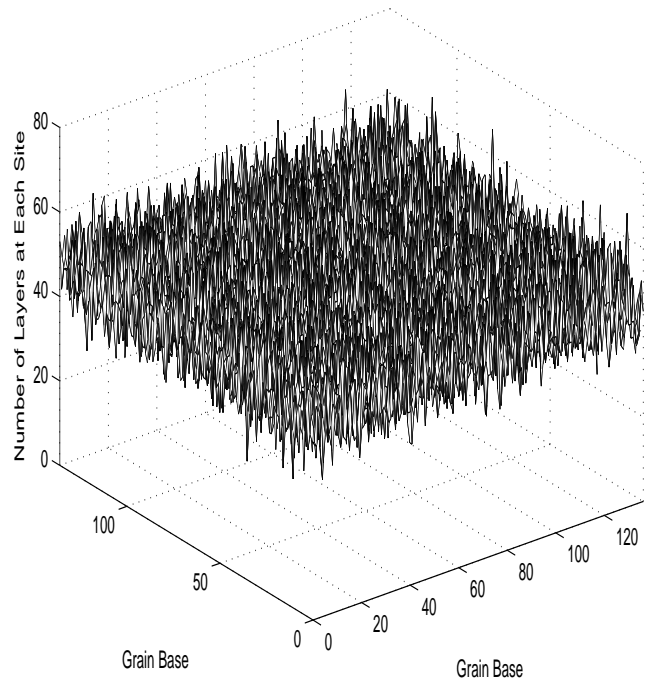}}
\caption{ Number of layers at each site on the grain surface after $2\times10^5$ yr of evolution at 10 K for Model 1.
  }
\label{fig6}
\end{figure}

\clearpage

\begin{figure}
\centering
\resizebox{12cm}{8cm}{\includegraphics{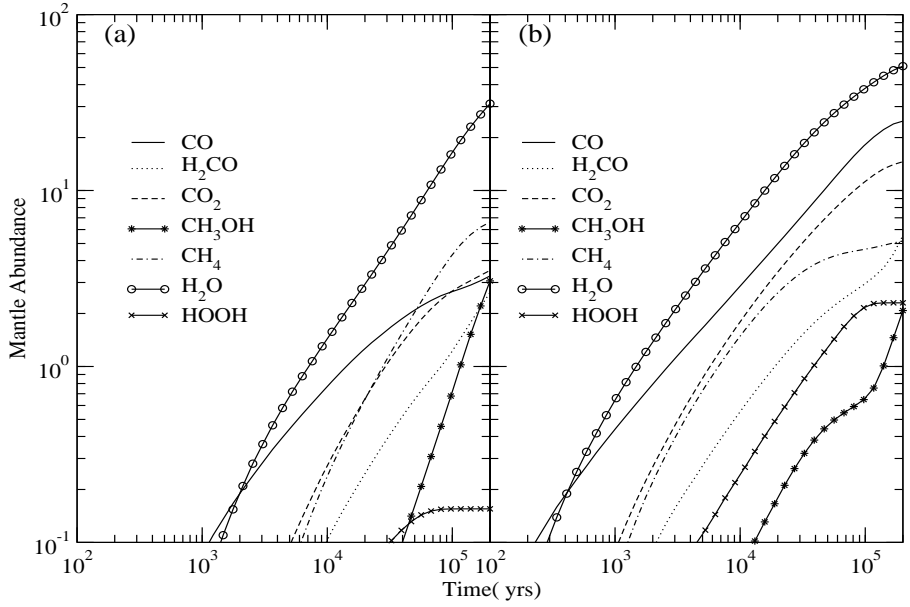}}
\caption{ The abundances of assorted species in 
the ice mantle are shown in terms of monolayers as a function of time at 10 K using Model 3 (Panel (a)) and Model 4 (Panel (b)).
}
\label{fig7}
\end{figure}

\clearpage

\begin{figure}
\centering
\resizebox{12cm}{8cm}{\includegraphics{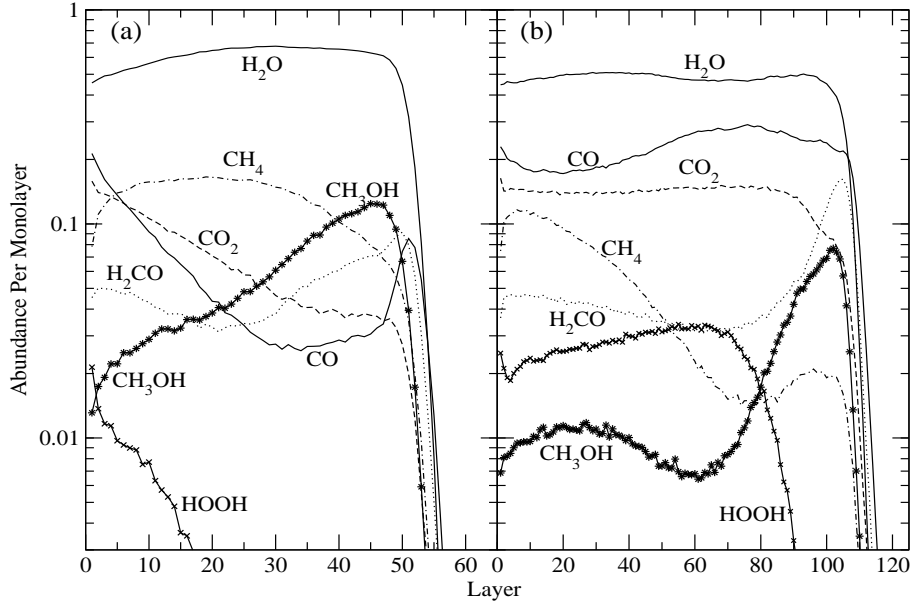}}
\caption{ The fraction of each monolayer occupied by  water, hydrogen peroxide, 
and major carbon-bearing species using Models 3 (Panel (a)) and 4 (Panel (b)) at 10 K shown as a function of monolayer at the final time of $2 \times 10^{5}$~yr. 
}
\label{fig8}
\end{figure}


\begin{table}
\caption{Abundances of Major Mantle Species at  10 K}
\label{table8}
\begin{tabular}{ c c c }
\hline
Species   & Model 3  & Model 4 \\ \hline
CO        &  10.5        &  48.7      \\
H$_2$CO   &  8.5       &   10.9     \\
CO$_2$    &  11.3       &   28.6    \\
CH$_3$OH  &  9.8        &   4.1   \\
CH$_4$    &  21.2       &   10.1   \\
HOOH      &  0.50       &   4.5   \\
H$_2$O    &  5.9(-5)    & 9.6(-5)  \\ \hline
\end{tabular}
\tablecomments{The abundances pertain to a time of $2\times 10^5$~yr.}
\tablecomments{The abundance of water ice is the fractional abundance with respect to gas-phase $n_{\rm H}$, while the abundances of the other ice components are percentages with respect to the water ice.   a(-b) means a$\times 10^{-b}$.
}
\end{table}

\begin{deluxetable}{l c c c c c}
\tablecaption{Comparison of Observational and Theoretical Ice Abundances}
\tablewidth{0pt}
  \tablehead{       &  CO  & CO$_2$ &  CH$_3$OH & CH$_4$ & HOOH}
  \startdata
  \hline
  Our Models & & & & \\
  \hline
Model 1, 10 K     &           5.7 & 19.6 & 27.7 & 18.5 & 4.8 \\            
Model 1, 13.5 K   &           8.4 & 13.4 & 17.3 & 20.6 & 0.77\\            
Model 1, 15 K     &           12.8 & 15.2 & 10.7 & 20.9 & 0.44 \\            
Model 1, 16.5 K   &           26.8 & 21.4 & 5.4 & 15.0 & 0.66 \\            
Model 2, 10 K     &           34.9 & 77.3 & 27.2 & 15.5 & 19.0 \\
Model 2, 13.5 K   &           48.1 & 38.7 & 8.9 & 12.0 & 7.1 \\
Model 2, 15 K     &           53.5 & 29.5 & 5.0 & 10.7 & 4.8 \\
Model 2, 16.5 K   &           58.5 & 26.6 & 2.4 & 8.4 & 2.8 \\
Model 3, 10 K     &           10.5 & 11.3 & 9.8 & 21.2 & 0.50\\
Model 4, 10 K     &           48.7  & 28.6 & 4.1 & 10.4 & 4.5\\ 
\hline
Previous Models & & & & \\
\hline           
Master equation, 10 K  &      2.6(-6) & 0.025 & 35.7 & 20.1 & \\
Master equation, 20 K  &      30     & 0.05  &  13  & 1.4 & \\
RE, slow diffusion 10 K  &    27.9 & 8.4(-14) & 8.4(-10) & 23.7 & 4.1(-6) \\   
RE, slow diffusion, 15 K &    57.5 & 1.3(-7) & 5.1(-5) & 5.6 & 5.7(-3)\\
RE, fast diffusion, 10 K &    36.3 & 2.4(-7) &  0.63 &  14.3 & 1.8(-8)\\
MRE, slow diffusion 10 K &   29.0 &  8.6(-13) & 1.3(-9) & 24.7 & 7.3(-6) \\
MRE, slow diffusion, 15 K &  54.8 &  1.4(-5) & 1.1(-4) & 7.2 & 3.9(-3)\\
MRE, fast diffusion, 10 K &  31.1 & 1.9(-6) & 2.0 & 14.3 & 3.4(-8)\\
Three phase model, 10 K &   10 & 5.7 & 1.1 & 30 &   \\
\hline
Observations & & & &\\
\hline
Toward Low Mass Stars  &    29 & 29 & 3 & 5 & \\ 
Toward High Mass Stars &    13 & 13 & 4 & 2 & \\
Toward Background Stars &    31 & 38 & 4 &  & \\
Elias 29               &        &    &   &  & $\leq$13 \\
Mon R2/IRS3            &        &    &   &  & $\leq$5\\
Mon R2/IRS2            &        &    &   &  & $\leq$13 \\
W51/IRS2               &        &    &   &  & $\leq$16 \\
\enddata
\label{table9}
\tablecomments{The abundances of the ice components are percentages with respect to the water ice.
  a(-b) means a$\times 10^{-b}$.  Observational results for carbon-bearing species are from the compilation of~\citet{Oberg2011},
 observational results for hydrogen peroxide are from ~\citet{Smith2011},
master equation results are from~\citet{Stantcheva2004}, while three-phase model results are from ~\citet{Garrod2011}}.
\end{deluxetable}

\begin{figure}
\centering
\resizebox{12cm}{8cm}{\includegraphics{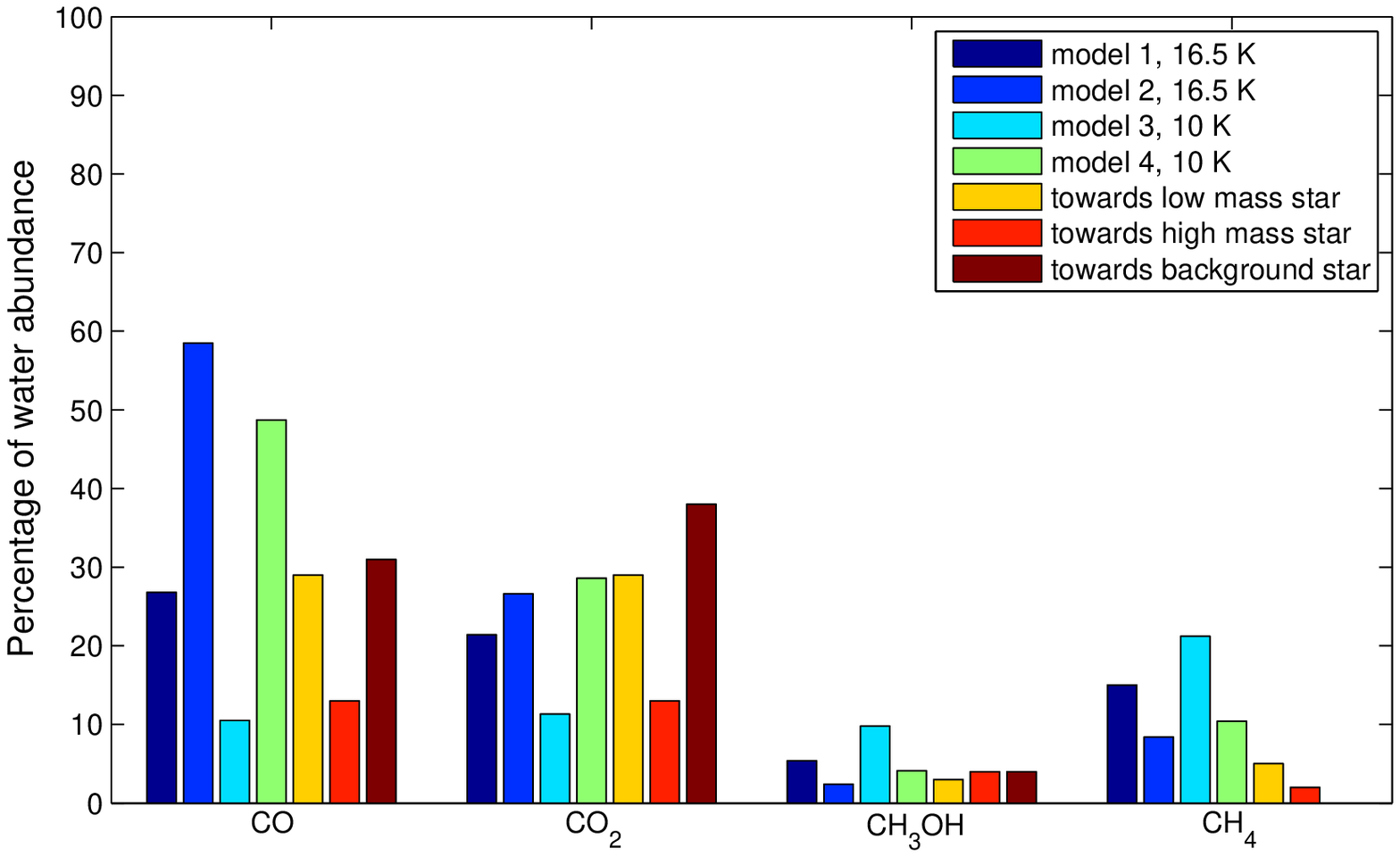}}
\caption{Histograms of the abundances of major carbon-bearing surface species by simulation and observation with respect to water ice.  Observational results are from 
~\citet{Oberg2011}. Model 1 and Model 2 results are at 16.5 K while Model 3 and Model 4 results are at 10 K. 
}
\label{fig9}
\end{figure}

The median fractional abundance of water ice in dense cold sources with respect to $n_{\rm H}$ is about $5\times10^{-5}$ 
\citep{Gibb2004,Pontoppidan2004,Boogert2004,Oberg2011}, which is in general agreement with our results for all models.
A compilation of Spitzer and ISO  observations by \citet{Oberg2011}
shows  that the median abundances of CO, CO$_2$, CH$_3$OH and CH$_4$ ice respectively as percentages of water ice are 29, 29, 3, 5  toward low mass stars, 
13, 13, 4, 2 toward high mass stars, and  31, 38, 4, - towards background stars, respectively.   
In Table \ref{table9}, we summarize the ice abundances from the observations of~\citet{Oberg2011} along with the results of assorted simulations in this paper and others.  Our simulations differ in their agreement with the observed results, although it is difficult to use the comparison to constrain both the temperature and the density.   In general, our methane results are significantly too high, while simulation results for the other three carbon species are much more reasonable.  For example,  values of \citet{Oberg2011} towards low-mass stars are quite reasonably fit by our Model 1 and Model 2 results at 16.5 K  while the observations towards high-mass stars are fit reasonably by Model 3 results at 10 K, with the exception of methane. The temperature dependence is non-intuitive because one might expect that the ices in front of high-mass and luminous 
sources possess a higher temperature than ices in front of low-mass and background stars.  Significantly higher ice temperatures would lead 
to less of the volatile species, such as CO and CH$_{4}$,  in the form of ices \citep{Oberg2011}.  Our models, on the other hand, 
associate a low CO ice abundance with a lower temperature, at which CO can be converted chemically to methanol.  
One must remember that with the fixed temperatures used in our model, we cannot distinguish between the 
temperature at which ice species form and the temperature, 
which may differ, at which they are observed. Thus, the predicted ice abundances 
in our model are most appropriately compared with observations with background stars.
The general degree of agreement can be seen in the histogram displayed in Figure~\ref{fig9}, where we have also included Model 4 at 10 K.  

We can also compare our results with respect to water ice with the results of other simulations, as shown in Table~\ref{table9}.  Using an approximate master equation stochastic approach over a time of 10$^{7}$ yr,  \citet{Stantcheva2004} calculated the abundances  of CO, CO$_2$, CH$_3$OH 
and CH$_4$ ice as a percentage of water ice.  Their reported results at 10~K and 20~K at a time of $3.2\times10^5$ yr  show very low abundances of CO$_{2}$ ice at both temperatures, and almost no CO at all at 10 K.  Instead, the conversion to methanol is much too efficient because they used an ultra-fast diffusion rate ($E_{\rm b}$/$E_{\rm D}$ = 0.3).  On the positive side, the master  equation results do show a low methane abundance at 20 K.
We also used our network to calculate results with both rate equation (RE) and modified rate equation (MRE) approaches to the surface chemistry with a so-called ``two-phase'' approach.  In this approach,  no distinction is made between the surface and interior monolayers.
The  RE and MRE simulations were undertaken at a density of $2 \times 10^{4}$ cm$^{-3}$ for both slow and fast diffusion.   We see a strong similarity between the results at $2 \times 10^{5}$ yr and the results of the master equation approach, where
there is hardly any  CO$_{2}$ formed.   Moreover, 
the production of CH$_3$OH is not efficient enough except for the MRE model with fast diffusion ($E_{\rm b}$/$E_{\rm D}$ = 0.5) at 10 K.  
Recently, three-phase models, in which the surface is distinguished from the interior monolayers, 
have been undertaken with the rate equation GRAINOBLE Model \citep{Taquet2012}, modified rate equations ~\citep{Garrod2011}
and macroscopic Monte Carlo approaches~\citep{Charnley2009}. Here, we list the results for the simulation by \citet{Garrod2011}, which partially includes stochastic effects, 
at 10 K and $2\times10^5$ yr.   This simulation is more successful except for the high abundance of methane.
Both the three-phase model of \citet{Garrod2011} and our simulations achieve a reasonable abundance of CO$_{2}$ at least partially because of the ``chain reaction'' mechanism discussed earlier.   Our CO$_{2}$ calculated abundances are in somewhat better agreement with the observation of \citet{Oberg2011}.

Recent analysis of spectroscopic data from different sources gives  upper limits to the HOOH abundance relative to water ice~\citep{Smith2011}. 
We can see in Table~\ref{table9} that most of our simulation results are below these upper limits, although there is far more HOOH produced in our models than 
 by rate equation or modified rate equation approaches. On the other hand, recent observational studies show that the fractional abundance of gas
phase HOOH toward $\rho$ Oph is about $10^{-10}$~\citep{Parise2011} while in Model 1 at 10 K, the gas phase HOOH fractional 
abundance is about $4\times 10^{-10}$. This shows that
moderate coverage of HOOH, about 1\% of the water abundance, is necessary to explain the gas phase abundance of HOOH unless a more efficient desorption mechanism
exists in the system.

It is also interesting to compare the distribution among the monolayers of major carbon-bearing species obtained from our simulation at $2 \times 10^{5}$ yr with inferences from observations and laboratory work \citep{Oberg2011}, by which one can distinguish among environments \citep{Tielens1991}.
Although the  situation is rather complex, \citet{Oberg2011} state that ``In general, observed ice features can, however, be explained by an early ice chemistry phase characteristic of hydrogenation of atoms, followed by atom-addition reactions in a CO-dominated ice phase.''   Moreover, it appears that the monolayer distributions of CO and CO$_{2}$ are anti-correlated; the fraction of CO found in a CO$_{2}$ environment and the fraction of CO$_{2}$ found in a CO environment are not high.  The normal interpretation is that the CO formed early via accretion and surface reactions is mainly converted into CO$_{2}$ (probably via CO + OH).   This conversion becomes inefficient at later times, where, for the low mass YSO case, there can be a high abundance of almost pure CO, with some conversion to formaldehyde and methanol.   To compare these inferences with our theoretical results, we focus on the final time of our calculations ($ 2 \times 10^{5}$ yr) at which time the water ice abundance has become large enough to agree with observations.  

Rather than trying to use our detailed figures, we divide the number of monolayers of ice into three equal groups - inner, middle, and outer, and list in Table~\ref{table10} the fraction of the monolayers in each group occupied by assorted molecular ices.  Of the different model results in this table, we see that Model 2 at 15 K appears to agree with the observational data best.  In particular, methane decreases from a fractional occupancy of 0.096 in the inner monolayers to an occupancy of 0.024 in the outer monolayers, while CO$_{2}$ decreases more weakly from 0.15 (inner monolayers) to 0.13 (outer monolayers).  On the other hand, both CO and its hydrogenation product, CH$_{3}$OH, increase in abundance from inner to outer monolayers.  Thus, we conclude that methane and carbon dioxide tend to be produced preferentially at early times, that CO and CH$_{3}$OH tend to be produced preferentially at later times, and that CO and CO$_{2}$  are in a sense anti-correlated spatially.   Other models fit the observational inferences more poorly, although Model 4 at 10 K is also reasonably good.   If we wish to divide environments into those that are polar and those that are non-polar, we can compare the sum of the abundances of non-polar (or relatively non-polar ) species -- CO, CO$_{2}$, O$_{2}$ --- with the sum of the abundances of polar species -- H$_{2}$O, HOOH, H$_{2}$CO, and CH$_{3}$OH.  If we do this, we find that there is no environment  in which the non-polar species are in the strong majority.  If there is a decidedly non-polar environment,  it would  have to occur within a smaller group of monolayers than those groups considered here.  Such an environment appears to occur at very high monolayers in Model 2 at 15 K and Model 4 at 10 K, as shown in Figures~\ref{fig5} and \ref{fig8}, where it can be seen that the CO and water abundances cross near the uppermost monolayers.
Nevertheless, it is fair to state that the model does not really reproduce the existence of a non-polar environment  where CO freeze-out is dominant.

\begin{deluxetable}{lrrrrrrrr}
\tablecaption{Distribution of Abundances into Inner (I), Middle (M), and Outer (O) Environments}
\tablewidth{0pt}
\tablehead{ Layers & CO & H$_{2}$CO & CO$_2$ & CH$_3$OH & CH$_4$ & H$_2$O & HOOH & O$_2$ }
\startdata
 Model 1, 10 K  &              &          &                  &          & & &     \\
I  & 0.055 &ÊÊ 0.037ÊÊ& 0.12ÊÊ&Ê  0.11 Ê Ê& ÊÊ 0.13ÊÊ&ÊÊ 0.46 & 0.045 & 0.026 \\
M &  0.012Ê& Ê 0.015ÊÊ&  0.092Ê&  0.17ÊÊÊ&ÊÊÊ 0.089ÊÊ & 0.60 & 0.013 & 0.0056\\
O &  0.012Ê & Ê 0.017ÊÊ&  0.088Ê & 0.20ÊÊ& ÊÊÊÊÊ 0.053Ê& Ê 0.59 & 0.0071 & 0.0037\\
Model 1, 15 K &                &          &                  &            & & &    \\
I & 0.12 & ÊÊ 0.031 & ÊÊÊ 0.12 &ÊÊÊ 0.030 &ÊÊÊÊÊ 0.14 &ÊÊ 0.55 & 0.0068 & 0.0030 \\
M &  0.051 &Ê 0.020 &ÊÊÊÊ 0.082 &Ê 0.058Ê &ÊÊÊÊ 0.14 &ÊÊÊ 0.64 & 0.00090 & 0.00032 \\
O & 0.064 &Ê 0.026ÊÊ&ÊÊ 0.071 &Ê 0.11 &ÊÊÊÊÊÊÊ 0.091Ê &  0.63 & 0.000020 & 0.000011\\
Model 3, 10 K &               &           &                   &             &  & &   \\
I & 0.10ÊÊ&Ê 0.043ÊÊ&ÊÊ 0.11 Ê& Ê 0.029ÊÊÊ&Ê 0.15ÊÊ&Ê 0.56 & 0.0074 & 0.0034 \\
M & 0.030Ê & 0.044ÊÊ&ÊÊ 0.049Ê&Ê 0.067Ê&ÊÊÊ 0.14ÊÊ&Ê 0.67 & 0.00032 & 0.00011 \\
O & 0.062Ê & 0.088Ê &ÊÊÊ 0.038 &ÊÊ 0.11Ê &ÊÊÊÊÊ 0.078Ê &  0.63 & 0 & 0 \\
Model 2, 10 K &                 &           &                  &           &  & &    \\
ÊI & 0.10ÊÊÊ&Ê 0.054ÊÊÊ&Ê 0.18ÊÊ&Ê 0.069ÊÊ&Ê 0.084Ê&Ê 0.34 & 0.067 & 0.055 \\
M & 0.12ÊÊÊ&Ê 0.072ÊÊ&ÊÊ 0.28ÊÊ&ÊÊ 0.070ÊÊ&Ê 0.027ÊÊ& 0.28 & 0.061 & 0.064\\
O & 0.10ÊÊÊÊ&  0.096ÊÊ&ÊÊ 0.27ÊÊ&ÊÊ 0.15ÊÊÊ&ÊÊ 0.012 &Ê 0.28 & 0.032 & 0.028 \\
Model 2, 15 K   &               &            &                &              & & &  \\
I & 0.19ÊÊ&ÊÊ 0.031ÊÊ&ÊÊ 0.15ÊÊ&Ê 0.011ÊÊ&ÊÊ 0.096ÊÊ&Ê 0.49 & 0.022 & 0.011 \\
M & 0.28ÊÊÊ&Ê 0.021ÊÊ&Ê 0.14Ê&ÊÊ0.0056Ê& 0.025Ê&ÊÊ 0.47 & 0.035 & 0.022 \\
O & 0.29ÊÊÊ& Ê 0.047ÊÊÊ&Ê 0.13ÊÊ&Ê 0.061ÊÊ&ÊÊ 0.024Ê&ÊÊ 0.44  & 0.0076 & 0.0040\\
Model 4, 10 K &      &             &                     &             &  & & \\
I & 0.19Ê&ÊÊÊ 0.043ÊÊ&ÊÊ 0.14ÊÊ&Ê 0.010ÊÊÊ&Ê 0.090Ê&Ê 0.49 & 0.025 & 0.013  \\
M & 0.26ÊÊ&ÊÊ 0.034ÊÊ&ÊÊ 0.15ÊÊ&Ê 0.0088ÊÊ& 0.027Ê&Ê 0.48 & 0.030 & 0.017 \\
O &  0.26Ê&ÊÊÊ 0.092ÊÊ&ÊÊ 0.11Ê&ÊÊ 0.049ÊÊÊ&Ê 0.018ÊÊ& 0.46 & 0.0037 & 0.0017 \\
\enddata
\label{table10}
\end{deluxetable}

\clearpage


\clearpage

\section{Discussion and Conclusions}
For the first time, we have performed a unified microscopic-macroscopic Monte Carlo simulation of gas-grain chemical modeling of a cold dense cloud.   In this approach, the
gas phase chemistry, which consists of more than 4000 reactions, is modeled by a  macroscopic Monte Carlo simulation,
while the surface/mantle chemistry, which consists of 29 reactions, is modeled by a microscopic Monte Carlo simulation, 
in which the position of each individual species is followed. For simplicity, we only allow H, O, C, and CO  to 
accrete onto grain surfaces. Although our surface reaction network is strongly simplified compared with a full surface reaction network, we can produce both water and the major carbon-bearing species on dust grains for a cold interstellar core. 

The unified simulation is made possible by modeling gas phase reactions as inhomogeneous Poisson processes and adopting the ``next reaction''
simulation method. In this approach, each chemical reaction in the gas phase has an absolute time for the reaction to start, which will change due to
the occurrences of other chemical reactions in the gas phase that change the abundances of its reactants. 
The advantage of modeling gas phase reactions  in this manner is that any desorption 
processes from grain surfaces can be treated in the same way as gas phase chemical reactions, thus making  it facile to couple gas phase processes with grain surface processes. 
This approach is superior to Gillespie's algorithm \citep{Gillespie1976}, which models gas phase reactions 
as homogeneous Poisson processes, in which abundances are not changed by other processes.
Our current model is superior to our earlier model \citep{Chang2007},  which couples a microscopic Monte Carlo 
simulation for the grain surface with rate equations for the gas,  because the restrictions required  
in our earlier model have been removed. 
In addition, most of the CPU time used in both approaches is spent on the microscopic Monte Carlo simulation of surface chemistry, so our new simulation does not lose much efficiency although the gas-phase chemistry is simulated by a Monte Carlo method instead of the more efficient rate equation approach.

Our results for gas-phase abundances show few differences from results obtained from rate equations, which means that they are in good agreement with observation at the time of $2 \times 10^{5}$~yr used for the surface chemistry. This result is not surprising; \citet{Stantcheva2004} showed that differences can show up only at longer times, where results on surfaces begin to affect gas-phase results more strongly.   Although our network does not contain surface reactions to produce  complex organic molecules in cold cores, as recently detected in the gas phase by \citet{bacmann2012}, we do produce gaseous methanol (CH$_{3}$OH) by photodesorption from the ice.  With Model 1 at 10 K,  we obtain a fractional abundance for methanol in the gas of $\approx 6 \times 10^{-10}$ at the final time of the calculation, which agrees well with the observed abundance in TMC-1 of $\approx 10^{-9}$ \citep{smith2004}.

Our results for the icy mantles can be compared with observations of total mantle abundances of assorted ices as well as their local environments in the ice.  Regarding total mantle abundances, our results show that after a time of $2 \times 10^{5}$ yr,  a number of models with different temperatures, densities and diffusion barrier-to-desorption energies  can reasonably reproduce the observed abundance of water ice as well as the main carbon-bearing species in front of high-mass YSO's, low-mass YSO's, and background stars, with the exception of methane, which is significantly overproduced \citep{Oberg2011}.  Regarding local environments,  if we divide the total number of monolayers into three groups (inner, middle, and outer) we can reproduce aspects of the inferred monolayer distributions and correlations for various species, but there is no one model that fits all of the information inferred from observations with high mass, low mass, and background star sources.


Although our model can successfully couple the time-dependent evolution of gas phase chemistry and grain surface chemistry, it is still not complete. First,
the surface reaction network is still too small. A complete surface reaction network typically consists 
of at least a few hundred reactions; however,  it is not clear if our model can treat this more complicated 
case without more advanced computer techniques and/or approximations.   Furthermore, to model sources where there is a change in the physical conditions as the chemistry occurs, such as a collapse or a warm-up, new algorithms will have to be developed to couple these changes into the unified Monte Carlo approach.

\acknowledgements

EH wishes to  acknowledge the support
of the National Science Foundation for his astrochemistry program. He also acknowledges support from the NASA Exobiology and Evolutionary Biology program through a subcontract from Rensselaer Polytechnic Institute.
We thank Anton I. Vasyunin and Karin {\"O}berg for helpful discussions.

\end{document}